\documentclass[12pt]{iopart}

\usepackage{iopams}
\usepackage{cite}
\usepackage{url}
\usepackage{color}

\usepackage{graphicx}
\usepackage{epstopdf}

\usepackage{amssymb}
\usepackage{color}

\usepackage{multirow}
\usepackage{graphicx}
\usepackage{hyperref}
\usepackage{graphicx}
\usepackage[normalem]{ulem}
\usepackage{color}
\usepackage{enumitem}
\usepackage{fancyhdr}
\usepackage{braket}

\newcommand \be{\begin{eqnarray}}
\newcommand \ee{\end{eqnarray}}
\newcommand{\del}{\partial}

\newcommand{\eqref}[1]{(\ref{#1})}
\def\Tr{\mathrm{Tr}}

\def\dd{\mathrm{d}}

\def\simge{\mathrel{
    \rlap{\raise 0.511ex \hbox{$>$}}{\lower 0.511ex \hbox{$\sim$}}}}
\def\simle{\mathrel{
    \rlap{\raise 0.511ex \hbox{$<$}}{\lower 0.511ex \hbox{$\sim$}}}}

\begin{document}


\title[]{Ornstein-Uhlenbeck diffusion of hermitian and non-hermitian matrices -- unexpected links}
\author{Jean-Paul Blaizot}
\address{IPTh, CEA-Saclay, CNRS/UMR 3681, F- 91191 Gif-sur-Yvette cedex, France}

\author{Jacek Grela}
\address{Marian Smoluchowski Institute of Physics, Jagiellonian University, ul. {\L}ojasiewicza 11, 30--348 Krak\'ow, Poland}
\author{Maciej A. Nowak$^*$}
\address{Marian Smoluchowski Institute of Physics and Mark Kac Complex Systems Research Center,  ul. {\L}ojasiewicza 11, 30--348 Krak\'ow, Poland}
\address{$^*$ Contributing author: maciej.a.nowak@uj.edu.pl}
\author{Wojciech Tarnowski}
\address{Marian Smoluchowski Institute of Physics, Jagiellonian University, ul. {\L}ojasiewicza 11, 30--348 Krak\'ow, Poland }
\author{Piotr Warcho\l{}}
\address{Marian Smoluchowski Institute of Physics, Jagiellonian University, ul. {\L}ojasiewicza 11, 30--348 Krak\'ow, Poland }
%

\begin{abstract}
We compare the  Ornstein-Uhlenbeck process for the Gaussian Unitary Ensemble to its non-hermitian counterpart - for the complex Ginibre ensemble.   
We exploit the mathematical framework based  on the generalized Green's functions,  which involves a new, hidden complex variable,  in comparison to the standard treatment of the resolvents. This new variable  turns out to be crucial to   understand  the pattern of the  evolution of  non-hermitian systems. The new feature is  the  emergence of the coupling between the  flow of eigenvalues and that of left/right eigenvectors. We analyze  local and global equilibria for both systems.  Finally, we highlight some unexpected links between both ensembles.  


\end{abstract}

\maketitle
{\bf SUBJECT AREA}: mathematical physics, theory of complex systems, random matrix theory\\
\section{Introduction}
In 1962, Dyson suggested an inspiring way to understand the joint probability distribution function (hereafter jpdf) of the eigenvalues of random matrices.  In order to find it, he introduced an auxiliary dynamics  in some  fictitious ``time", which, in the large time limit,  lead to a stationary state (Gibbs state) representing the desired jpdf. 
As he pointed~\cite{DYSON}, {\it ``after considerable and fruitless efforts to develop a Newtonian theory of ensembles, we discovered that the correct procedure is quite different and much simpler. The $x_i$} [eigenvalues] {\it  should be interpreted as positions of particles in Brownian motion"}. The resulting stationary distributions (originally for hermitian or for unitary random matrices) were obtained as a result  of Ornstein-Uhlenbeck diffusion with a drift force coming from electrostatic-like repulsion of eigenvalues. The success of this description has contributed  to multiple applications of random matrix models in practically all branches of science. The notion of ``time" has evolved as well, so nowadays it can be a physical parameter, representing either the real time or, e.g., the length of a mesoscopic wire, the area of a string or an external temperature.  The idea of a noisy walk of eigenvalues recently led also to such concepts as  determinantal processes~\cite{FORRESTER,KOBIZKA,SCHEHR}, Loewner diffusion~\cite{LOEWDIFF}, fluctuations of non-intersecting interfaces in thermal equilibrium~\cite{NADALMAJUMDAR} and the emergence of pre-shock spectral waves and universal scaling at the critical points of several random matrix models. 
 
Three years after Dyson, Ginibre~\cite{GINIBRE} has considered for the first time strictly non-hermitian random matrix models, whose spectrum does not need to be confined either to the real line (hermitian operators) or to the unit circle (unitary operators), but can be located on a two-dimensional support on the complex plane. The original motivation for the study of complex, random spectra was purely academic. Today however, non-hermitian random operators play a role in quantum information processing, in financial engineering (when lagged correlations are discussed~\cite{BIELYTHURNER}) or in identifying clusters in social or biological networks using non-backtracking operators~\cite{KRZAKALA}, to name just a few recent applications.  Additionally, statistical properties of eigenvectors of non-hermitian operators contribute to understanding scattering problems in open chaotic cavities~\cite{PERSSONGORIN} and random lasing. 

It is surprising that in the last  half century, the Dysonian  picture of random walk  of eigenvalues was not applied to the complex Ginibre Ensemble (GE). The Brownian walk problem for the real Ginibre Ensemble was recently studied in \cite{ZABORONSKY}. In this contribution, partially based on our earlier work on this subject, we show how to fill this logical gap, and we also speculate on the reasons why the non-hermitian extension of a random walk scenario was far from obvious. 

In Section~2, we start from recalling  Dyson's original construction~\cite{DYSON}. Then, we propose an alternative description, where the fundamental object is the characteristic polynomial. We show the advantages of such  description, borrowing heavily from the analogies to the simplest model of turbulence, i.e., the so-called Burgers equation. We also briefly mention, how the  seminal results for the Gaussian Unitary Ensemble (GUE) can be recovered from a Burgers-like description. 

In Section~3, we formulate a mathematical framework, which allows us to parallel the turbulent picture in the case of the GE. In particular, we unravel a hidden dynamics associated with  a new complex variable, which in standard descriptions of non-hermitian random matrix models  is treated as an infinitesimal regulator only. We point out, that the non-hermitian  character of the GE binds the dynamics of eigenvalues to the evolution of eigenvectors in a non-trivial way. Alike in the case of GUE, we demonstrate how the well-known results of the GE can be easily reclaimed in our formalism. Section~4 contains numerical experiments for both GUE and GE capturing the relevant diffusive degrees of freedom. 
In Section~5, we uncover the unexpected links between the descriptions of the Gaussian Unitary and the Ginibre ensembles. Section~6 concludes the paper and lists some open problems on noise in matrix models. 

\section{Diffusion in the Gaussian Unitary Ensemble}

 According to Dyson, the eigenvalues of a random, $N$ by $N$ hermitian matrix belonging to the GUE fulfill the following stochastic equation
 \be
  {\rm d}\lambda_i(\tau)=\frac{1}{\sqrt{N}} {\rm d}B_i(\tau)+\frac{1}{N}\sum_{j=1, j\neq i}^N \frac{1}{\lambda_i-\lambda_j}  {\rm d}\tau -a \lambda_i  {\rm d}\tau,
  \label{LangGUE}
  \ee
  where $B_i$'s are one-dimensional standard Brownian motions, $\lambda_i$'s denote the eigenvalues and $\tau$ is the time variable. The second term represents a fictitious electric field coming from the logarithmic Coulomb potential (originating  from the Van der Monde determinant) and the last term represents the drift coming from the confining harmonic potential (Ornstein-Uhlenbeck process). In the limit when $N$ tends to infinity and $\tau \rightarrow \infty$, the eigenvalues freeze-out as a result of the compromise between the repulsion (electric field) and attraction (harmonic potential). The resulting spectral distribution takes the form of the Wigner semicircle. Despite the fact that Dyson was primarily interested  in the equilibrium state, and  introduced a time in an auxiliary construction, he pointed out that the transition to the equilibrium is quite subtle.  
In his own words~\cite{DYSON}, the Coulombic term is {\it ``measuring the frequency with which two charges come into coincidence. This term is mainly sensitive to the local (microscopic) configurations of the gas particles... at the microscopic  time scale ... After local equilibrium is established... the gas must adjust itself by macroscopic motion on the time scale"}, which is $N$ times larger compared to the microscopic one. He also noted that \cite{DYSON}{\it "a rigorous proof that this picture is accurate would require a much deeper mathematical analysis"}. The discussion in this section gives  support to this picture.

Let us introduce an $N\times N$ hermitian matrix $H$ by defining its complex entries according to:
\begin{eqnarray}
H_{ij}= \left\{ 
	\begin{array}{l r}
	x_{ii}, & i=j, \\
	x_{ij}+i y_{ij}, & i\neq j,
	\end{array}\right.
\end{eqnarray}
where $x_{ij}=x_{ji}$ and $y_{ij}=-y_{ji}$, with $x_{ij}$ and $y_{ij}$ real. Furthermore let $x_{ij}$ and $y_{ij}$ perform white noise driven, independent random walks, such that
\begin{eqnarray}
	\left \langle\delta H_{ij} \right\rangle = -a H_{ij} \delta \tau, \qquad \left \langle \left|\delta H_{ij}\right|^{2}\right\rangle=\frac{g_{ij}}{N}\delta \tau ,
	\label{hermincr}
\end{eqnarray} 
with $g_{ij} = 1+\delta_{ij}$ and for any $i$ and $j$. Let $P(x_{ij},\tau)P(y_{ij},\tau)$ be the probability that the off diagonal matrix entry $H_{ij}$  will change from its initial state to $x_{ij}+i y_{ij}$ after time $\tau$. Analogically, $P(x_{ii},\tau)$ is the probability of the diagonal entry $H_{ii}$ becoming equal to $x_{ii}$ at $\tau$.
The evolution of these functions is governed by the following Smoluchowski-Fokker-Planck (SFP) equations:
\begin{eqnarray}
\label{entrydiff}
\frac{\del}{\del \tau}P(x_{ii},\tau) & = \left(\frac{1}{2N}\frac{\del^{2}}{\del x_{ii}^{2}}+a\frac{\del}{\del x_{ii}}x_{ii} \right)P(x_{ii},\tau) , \nonumber \\
\frac{\del}{\del \tau}P(v_{ij},\tau) & = \left(\frac{1}{4N}\frac{\del^{2}}{\del v_{ij}^{2}}+a\frac{\del}{\del v_{ij}}v_{ij} \right)P(v_{ij},\tau) , \quad i<j, 
\end{eqnarray}
where the parameter $a$ measures the strength of the harmonic potential confining the diffusion of the matrix elements and $v_{ij}$ denotes either $x_{ij}$ or $y_{ij}$. The joint probability density function is thus defined as
\begin{eqnarray}
	 \qquad P(x,y,\tau)\equiv\prod_{k=1}^NP(x_{kk},\tau)\prod_{i<j=1}^N P(x_{ij},\tau)P(y_{ij},\tau) \label{prob}
\end{eqnarray}
and satisfies the following equation
\begin{eqnarray}
\label{jpdfeq}
\del_\tau P(x,y,\tau)=\mathcal{A}(x,y) P(x,y,\tau),
\end{eqnarray}
with
\begin{eqnarray}
 \mathcal{A}(x,y) & = \sum_{k=1}^N \left(\frac{1}{2N}\frac{\del^{2}}{\del x_{kk}^{2}}  +a\frac{\del}{\del x_{kk}}x_{kk}\right) +\frac{1}{4N}\sum_{i<j=1}^N \left ( \frac{\del^{2}}{\del x_{ij}^{2}}+ \frac{\del^{2}}{\del y_{ij}^{2}} \right ) + \nonumber \\ 
& + a\sum_{i<j=1}^N \left ( \frac{\del}{\del x_{ij}}x_{ij}+ \frac{\del}{\del
y_{ij}}y_{ij} \right ) . \label{prob1}
\end{eqnarray}
A source-like solution of \eqref{jpdfeq} reads:
\begin{eqnarray}
	\label{solgue}
	P(x,y,\tau) = C \exp \left ( - \frac{Na}{1-e^{-2a\tau}} \Tr (H - H_0 e^{-a\tau})^2) \right ) ,
\end{eqnarray} 
with $H(\tau=0)=H_0$ and $C$ is a normalization constant.
With the setting thus defined, let us proceed to the derivation of the partial differential equations obeyed by  the averaged characteristic polynomial (hereafter called ACP)
 $U(z,\tau)$  associated with the diffusing matrix $H$:
\be 
\label{defu}
U(z,\tau)\equiv\langle \det\left(z-H\right)\rangle_{\tau},
\ee
where the angular brackets denote the averaging over the time dependent probability density (\ref{prob}). 
In Appendix A, we show that the ACP satisfies
\begin{eqnarray}
\del_{\tau}U(z,\tau)=-\frac{1}{2N}\del_{zz}U(z,\tau)+a z \del_{z}U(z,\tau)- a N U(z,\tau).
\label{eq:U}
\end{eqnarray}
Note that the standard Green's function (resolvent) 
 associated with $H$  is related to ACP in the large $N$ limit
\be
G\equiv G( z, \tau)\equiv   \lim_{N\to\infty} \frac{1}{N}\langle {\rm Tr} \frac{1}{z-H} \rangle= \lim_{N\to\infty}\frac{1}{N} \del_{z} \ln U.
\ee
Thus, the spectral density of $H$ is given by
\be
\rho(\lambda,\tau)= -\frac{1}{\pi} \lim_{\epsilon \to 0_+}{\rm Im}G(z=\lambda+i\epsilon,\tau ),
\label{SPl}
\ee
through the Sokhotski-Plemelj formula.

Let us define $f_N\equiv f_N(z,\tau)\equiv \frac{1}{N} \del_{z} \ln U$ (this is the complex analogue of the Cole-Hopf transform).  Due to Eq.~(\ref{eq:U}),  $f_N$ satisfies:
\be
\del_{\tau} f_N+f\del_z f_N-a \del_z \left(z f_N \right)=-\frac{1}{2N}\del_{zz}f_N.
\ee
Since $\lim_{N \rightarrow \infty}f_N=G$,  we see that the Green's function is governed by the following complex Burgers-like differential equation:
\be
\del_{\tau} G+G\del_z G-a \del_z \left(z G \right)=0. 
\label{Burgers}
\ee
 \subsection{Macroscopic equilibrium} 
In the $\tau\to \infty$ limit the time derivative in Eq.~\eqref{Burgers} vanishes and so this equation is easily solvable since it reduces to: 
\be
\partial_z \left(\frac{1}{2}G^2 -azG\right)=0.
\label{Wig}
\ee
Because all moments $\frac{1}{N}\left <Tr H^k \right >$ are finite for any $k$,  the function $G(z,\tau)$  has to tend to zero as $1/z$ in the $z \rightarrow \infty$ limit. This observation fixes the integration constant of Eq.~(\ref{Wig}) to be equal to $-a$. The resulting solution of the quadratic equation
\be
\frac{1}{2}G(z)^2-azG(z)+a=0
\ee
reads $G(z)=a(z-\sqrt{z^2-2/a})$.
 Using (\ref{SPl}), for the standard value $a=1/2$, we recover the Wigner semicircle
 \be
 \rho(x)=\frac{1}{2\pi} \sqrt{ 4-x^2}.
 \ee
 The analogy to the Burgers equation is however deeper, as we pointed out in ~\cite{USOLD}. For the real Burgers equation, the solution based on the method of characteristics breaks down due to the emergence of the pre-shock wave (singularity). Similar phenomenon takes place on the complex plane for Eq.~(\ref{Burgers}). The resulting spectral shock waves correspond to the endpoints of the spectrum. Several other, surprising  links between the Burgers equation and some probabilistic model are discussed in~\cite{MENON}. 
\subsection{Microscopic limit} 
We stress here that the equation (\ref{eq:U}) for the diffusing characteristic polynomial is exact for any $N$ and for any initial conditions. We can therefore use it to retrieve the spectral features at all time scales and at all points of the spectrum. In order to simplify the analysis, let us start by performing a useful change of variables belonging to the class of Lamperti transformations~\cite{LAMPERTI1,LAMPERTI2,LAMPERTI3,CEPALEPINGLE}. This change of variables will allow to establish a connection between the Ornstein-Uhlenbeck process and the case of free diffusion ($a=0$), where the results are known~\cite{USOLD,USACTA}. 
To check that indeed we can recover a free diffusion equation from the Ornstein-Uhlenbeck process, we explicitly write down the relevant Lamperti transformation:
\begin{eqnarray}
	& U(z,\tau) = (1+2a\tau')^{-N/2} U'(z',\tau'), \nonumber \\
	& z' = e^{a\tau} z ,\quad \tau' = \frac{1}{2a} \left ( e^{2a\tau} - 1\right ).
	\label{lamperti}
\end{eqnarray}
Straightforward but lengthy calculations yield
 \begin{eqnarray*}
	& \partial_\tau U = A^{-2} \left ( - \frac{Na}{1+2a\tau'} BU' + B \partial_{\tau'} U' \right ) + a z' B\partial_{z'} U', \\
	& \partial_{z} U = BA^{-1} \partial_{z'}U', \qquad \partial_{zz} U = B A^{-2}\partial_{z'z'} U',
\end{eqnarray*}
where  we define $A = e^{-a\tau}$ and $B= \left ( 1 + 2 a \tau' \right )^{-N/2}$.
Plugging these into \eqref{eq:U} produce a free diffusion equation in the variables $(z',\tau')$:
\begin{eqnarray}
	\partial_{\tau'} U' = - \frac{1}{2N} \partial_{z'z'} U'.
	\label{hermdiff}
\end{eqnarray}
In order to avoid unnecessary repetitions, we highlight here only the consequences of Eq.~\eqref{hermdiff}, relegating the details to already published work~\cite{USOLD,USACTA}.
First, it is exactly integrable on the complex plane for any initial conditions. 
The corresponding Cole-Hopf transformation maps the diffusion equation onto the complex viscid Burgers equation
\be
\partial_{\tau^{'}}f_N^{'} +f^{'}\partial_{z^{'}}f_N^{'}=-\nu_s \partial_{z^{'}z^{'}}f_N^{'},
\ee
where the (negative) spectral viscosity reads $\nu_s=1/2N$. 
Second, in the large $N$ limit (inviscid limit), spectral shock waves form at the endpoints of the spectra. 
Third, in the vicinity of the shock waves (endpoints of the spectra), the above equation captures the microscopic universality of the GUE ensembles, leading to Airy function oscillations at the edges. Intuitively, spectral oscillations at the endpoint origin from the negative sign of viscosity,  which causes the ``roughening" of the transition  instead of smoothening it, as would be expected in the case of a positive viscosity. 
Last but not least, the above picture confirms that Dysonian microscopic equilibrium is formed already at very short time scales and its character is determined solely by the global properties of the random matrix, i.e. the symmetries  deciding on  the functional form of the Coulombic repulsion.   

\section{Diffusion in the Ginibre Ensemble}
\label{secge}
Contrary to the case of the hermitian ensembles, the spectrum of non-hermitian matrices is genuinely complex. Let us define the simplest example, the so-called complex Ginibre Ensemble where each element of the $N\times N$ matrix $X$ is drawn from  a complex Gaussian distribution. That is, each entry $X_{ij}=x_{ij}+iy_{ij}$ consists of 
$x_{ij}$ and $y_{ij}$ drawn from standard Gaussian distributions. Note that all moments $ \left <{\rm Tr} X^n \right >$ (and $\left <{\rm Tr}(X^{\dagger})^n  \right >$) vanish because $\left < \Tr X^2 \right > = 0$. The only non vanishing moments are the mixed ones, i.e. $\left < {\rm Tr} (XX^{\dagger})^n \right >$. As we will show, in the large $N$ limit, the eigenvalues condense uniformly on the centered disc on the complex plane. Therefore, the spectrum exhibits a jump at the rim, contrary to the hermitian cases, when the real spectrum is continuous at the endpoints, and only the derivatives of the  spectrum are discontinuous. Moreover, the spectrum is non-analytic inside the disc, which seems to disqualify all the methods based on analyticity of the complex variable $z$.  This is best visible, when we try to repeat the hermitian construction for the Green's function
$G(z)=\frac{1}{N}\left<{\rm Tr} \frac{1}{z-X}\right>$. Since all moments   vanish, such a Green's function is simply equal to $G(z)=\frac{1}{z}$, and does not reflect correctly the  spectral properties of the ensemble. Similarly, the characteristic determinant is trivial, $\left <\det (z-X) \right >=z^N$.

The way out, based on an electrostatic analogy, was suggested a long time ago~\cite{OLD}.
We define an electrostatic potential 
\be
\Phi \equiv \Phi(z)=\lim_{\epsilon \rightarrow 0} \lim_{N \rightarrow \infty} \frac{1}{N} \left\langle {\rm Tr} \ln [|z-X|^2 +\epsilon^2] \right\rangle\ ,
\label{Coulomb}
\ee
where we  use a short-hand notation:
$|z-X|^2 + \epsilon^2 = (z {\bf{1}}_N -X)(\bar{z}{\bf{1}}_N-X^{\dagger}) + \epsilon^2 {\bf{1}}_N$, where $\bf{1}_N$ is the $N$-dimensional identity matrix. Then,  we calculate  the ``electric field" as a  gradient of the electrostatic potential, 
\be
G(z,\bar{z})=\partial_z \Phi=\lim_{\epsilon \rightarrow 0} \lim_{N \rightarrow \infty} \frac{1}{N} \left< {\rm Tr} 
\frac{\bar{z}-X^{\dagger}}{|z-X|^2 +\epsilon^2} \right> .
\label{Electric}
\ee
The electric field plays the role of the correct Green's function. Indeed, applying the  Gauss law, in the next step,
\be
 \rho =\frac{1}{\pi}\partial_{\bar{z}}G = \frac{1}{\pi} \partial_{z\bar{z}} \Phi = \lim_{\epsilon \rightarrow 0} \lim_{N \rightarrow \infty} \frac{1}{\pi N} \left\langle {\rm Tr} \frac{\epsilon^2}{[|z-X|^2 +\epsilon^2]^2} \right\rangle
 \label{Gauss} ,
 \ee  
we recover the spectral density $\rho(z,\bar{z})=\frac{1}{N} 
\langle \sum_{i=1}^N \delta^{(2)} (z-\lambda_i) \rangle$, using the known representation of the two-dimensional delta function
$\delta^{(2)}(z)=\lim_{\epsilon \rightarrow 0} \frac{1}{\pi} \frac{\epsilon^2}{[|z|^2+\epsilon^2]^2}$.  Note that the Gauss law implies the non-analyticity of $G(z,\bar{z})$. 
It is crucial that the limit $N\rightarrow \infty$ is taken first, before taking the infinitesimal regulator $\epsilon$ to zero, since only such order provides the necessary coupling between $X$ and $X^{\dagger}$, reflected in non-vanishing mixed moments.   If one took the limits in an opposite order, $X$ and $X^{\dagger}$ would decouple, and we  would obtain a trivial result $G(z)=1/z$. The bad news, however, is that the Green's function $G(z,\bar{z})$ (\ref{Electric}) is given by a very complicated expression, without any similarity to the standard form of the resolvent. 

One may bypass the difficulty by relying  on the the algebraic construction for the so-called generalized Green's functions proposed some time ago~\cite{JANIKNOWAK,JAROSZNOWAK}.  First, we notice that
\be
\hspace{-1cm} {\rm Tr} \ln [|z-X|^2+\epsilon^2]= \ln \det [|z-X|^2+\epsilon^2] = \ln \det  \left( \begin{array}{cc}
z -X & i\epsilon \\ i\epsilon & \bar{z}-X^{\dagger} \end{array} \right) ,
\ee
where the argument of the last determinant is a $2N\times 2N$ matrix,  built out of four $N\times N$ blocks. 
Let us now define a new operation called a block-trace, defined as ${\rm bTr} \equiv {\bf{1}}_2 \otimes \Tr_{N\times N}$, which acts in the following way:
\be
{\rm bTr} \left( \begin{array}{cc}
A & B \\ C & D \end{array} \right)_{2N \times 2N}  \equiv \left( \begin{array}{cc}
{\rm Tr} A & {\rm Tr} B \\ {\rm Tr} C& {\rm Tr} D \end{array} \right)_{2\times 2},
\ee
converting a $2N\times 2N$ block matrix into a $2 \times 2$ matrix built out of ordinary traces. 
Additionally, we define another pair of block matrices
\be
Q= \left( \begin{array}{cc}
z & -\bar{w} \\ w & \bar{z} \end{array} \right) ,  \,\,\,\,\,\,\,\,{\cal{X}}=\left( \begin{array}{cc}
X & 0 \\ 0 & X^{\dagger} \end{array} \right).
\ee
We are now ready to propose the construction of the generalized resolvent ($2 \times 2$ matrix)
\be
{\cal{G}}(z,w) &\equiv& \left( \begin{array}{cc}
 {\cal{G}}_{11} & {\cal{G}}_{1\bar{1}} \\ {\cal{G}}_{\bar{1}1} & {\cal{G}}_{\bar{1}\bar{1}} \end{array} \right) =
 \frac{1}{N} \left< {\rm bTr}\frac{1}{Q-\cal{X} }\right>,
\label{genGreen}
 \ee
By construction, $\mathcal{G}_{11}$  is equal to the non-analytic resolvent $G(z,\bar{z})$ (\ref{Electric}), provided we identify 
$|w|^2=\epsilon^2$.  Note that  the duplication trick allowed us to linearize the problem,   
since  the form of the generalized resolvent (\ref{genGreen}) has formally  the form of the standard resolvent for hermitian matrices. 
One may ask the question,  what role is played by the three remaining elements of the matrix ${\cal G}$? 
Let us recall, that the general (non-normal) matrix $X$ is determined in terms of its eigenvalues ($Z$)  and a set of left ($\bra{L}$) and right ($\ket{R}$) eigenvectors  ($X=\sum_i z_i \ket{R_i} \bra{L_i}$),  which are bi-orthogonal $\braket{L_i|R_j} = \delta_{ij}$. By applying a transformation $S={\rm diag} (R, L)$, $S^{-1}={\rm diag } (L^{\dagger}, R^{\dagger})$
(where  $L,R,L^{\dagger},R^{\dagger}$ are $N\times N$ matrices built from the corresponding eigenvectors),  we notice that
\be
\det (Q-{\cal X})= \det [S^{-1}(Q-{\cal X})S]=\det \left( \begin{array}{cc}
z -Z & -\bar{w}L^{\dagger}L \\ wR^{\dagger}R & \bar{z} -Z^{\dagger} \end{array} \right) ,
\ee
so the off-diagonal  elements of the generalized Green's functions are related to the expectation values of the overlaps of eigenvectors. Indeed,
 the left-right eigenvector correlator~\cite{CHALKERMEHLIG} reads:
 \be O(z, \tau)\equiv \frac{1}{N^2} \left< \sum_a O_{aa} \delta^{(2)}(z-z_a) \right>,
 \label{eigencor}
 \ee
 where $O_{ij}=\left <L_i|L_j \right > \left <R_j|R_i \right >$ is given in the large $N$ limit by the product of off-diagonal elements of $\mathcal{G}$:
 \be
	\lim_{N\to \infty} O(z,\tau) = -  \frac{1}{\pi} {\cal G}_{1 \bar{1}} {\cal G}_{\bar{1}1} |_{w=0} .
\ee
as was proven in \cite{NOERENBERG}.
 The appearance of this correlator is a genuine feature of non-hermitian random matrix models, since in the hermitian case left and right eigenvectors coincide and so $O_{ij}=\delta_{ij}$. Finally, for completeness we notice that ${\cal G}_{\bar{1}\bar{1}}$ is a complex conjugate of ${\cal G}_{11}$ and does not bring any new information. 
 
\begin{figure}[ht!]
	\centering
	\includegraphics[width=.8\textwidth]{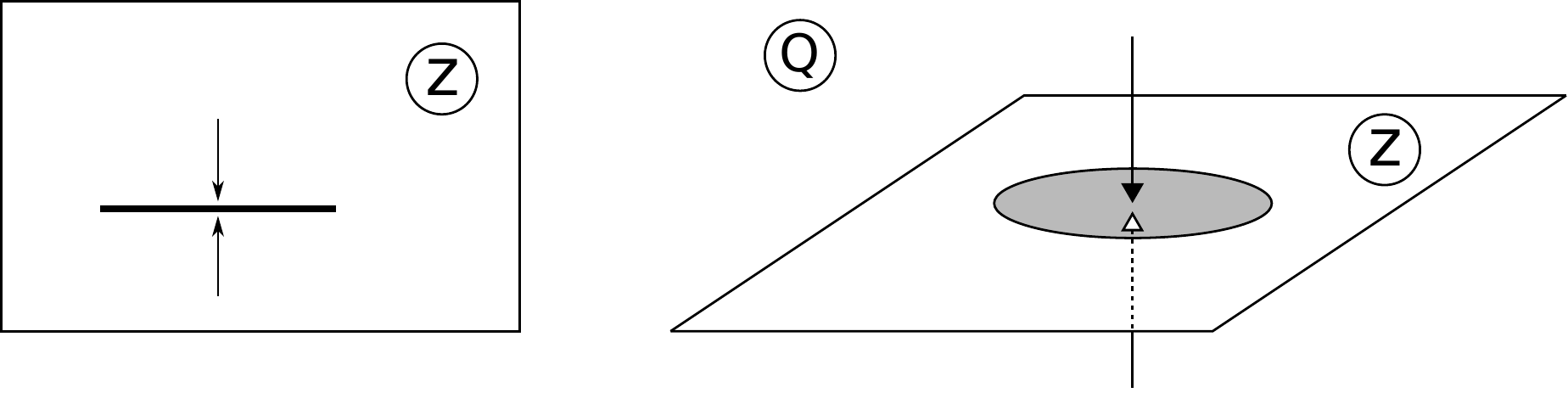}
      \caption{Schematic comparison between the domains of hermitian Green's function $G(z)$ and the non-hermitian, generalized Green's function ${\cal G}(Q)$.  Arrows on the left figure signal the discontinuity of the Green's function when approaching the cut (solid line), arrows on the right figure  denote an additional variable $w$, which in standard approach is treated as only an infinitesimal regulator. Shaded disc represents the non-analytic domain where the eigenvalues condense. }
      \label{Figurezw}
\end{figure}

We would now like to comment on the role of the $w$ variable. In the hermitian case, the method of the resolvent involves  the whole complex plane $z$,  despite the fact that the real spectrum comes only as a discontinuity near $z=\lambda \pm  i \epsilon$, corresponding to  the imaginary part of the resolvent.  In the non-hermitian case, the spectrum is complex, but one may be tempted to probe the generalized Green's function with the complex plane $w$ "orthogonal"  to the plane $z$, as schematically depicted on Fig.~\ref{Figurezw}. This choice of strategy is reinforced by the above observed coupling of the $w$ plane to eigenvector correlators. 

 The promotion of the original regulator $i\epsilon$ to a complex variable $w$ has additional advantages.    
 Firstly, from the algebraic point of view, $Q$ is a quaternion, since $Q=q_0 + i \sigma_jq_j$, where $\sigma_j$ are Pauli matrices, so  $z=q_0+iq_3$  and $w=-q_2+iq_1$. This  fact significantly simplifies the algebraic calculations, since block matrices such as $\mathcal{X}$ and arguments $Q$ naturally appear in non-hermitian random matrix models, e.g. in the generalized Green's function technique \cite{JANIKNOWAK,JAROSZNOWAK},in hermitization methods \cite{GIRKO,FEINBERGZEE,CHALKERWANG} or in the derivation of the multiplication law for non-hermitian random matrices \cite{BJN}. The above construction was also recently proven rigorously in the mathematical literature \cite{BELINSCHI}. 
Secondly, introducing the variable $w$ provides a hint on which object should play the role of the average characteristic polynomial in the case of the Ginibre ensemble subjected to Ornstein-Uhlenbeck process, as we now demonstrate. 

 We define now a determinant expressed in terms of the quaternionic variable $Q$ by. 
\be
D\equiv D(|w|, z, \bar{z}, \tau)\equiv  
\left\langle {\rm det} (Q -\cal{X}) \right \rangle_{\tau} 
=\left\langle \det\left( \begin{array}{cc} z-X & -\bar{w} \\ w & \bar{z}-X^{\dagger} \end{array}\right)\right\rangle_\tau ,
\label{cd}
\ee
 where the measure over which the averaging is performed reflects the Ornstein-Uhlenbeck process. More concretely, it is given by the following averaged increments (compare with the hermitian counterpart \eqref{hermincr}):
\be
\left < \delta X_{ij} \right> = - a X_{ij} \delta \tau, \qquad \left < \left | \delta X_{ij}\right |^2 \right > = \frac{1}{N} \delta \tau ,\label{nonhdiff}
\ee
which is also expressible as a Smoluchowski-Fokker-Planck equation for the jpdf $P(x,y,\tau) = \prod_{i,j=1}^N P(x_{ij},\tau) P(y_{ij},\tau)$: 
\begin{eqnarray}
\label{ggeq}
\del_\tau P(x,y,\tau)=\mathcal{B}(x,y) P(x,y,\tau),
\end{eqnarray}
with
\begin{eqnarray}
\mathcal{B} = \frac{1}{4N} \sum_{i,j=1}^N \left ( \partial^2_{x_{ij}} + \partial^2_{y_{ij}} \right ) +  a  \sum_{i,j=1}^N \left ( \partial_{x_{ij}} x_{ij} + \partial_{y_{ij}} y_{ij} \right).
\end{eqnarray}
Up to an irrelevant normalization constant $C$, a source-like solution to \eqref{ggeq} reads
\begin{eqnarray}
	P(X,\tau) = C \exp \left ( - \frac{2Na}{1-e^{-2a\tau}} \Tr |X - X_0 e^{-a\tau}|^2 \right ) .
\end{eqnarray}
Following similar steps as in the hermitian case, we express the determinant with the help of the auxiliary Grassmann variables and use the properties of the diffusion process  to arrive (cf. Appendix B for the details) at the exact (for any matrix size $N$ and for any initial conditions) equation 
 \begin{eqnarray}
 \label{ggdiff}
	\del_\tau D = \frac{1}{N} \del_{w\bar{w}} D - 2N  a  D +  a  \textrm{d} D ,
\end{eqnarray}
with the operator $\textrm{d} = z \partial_z+\bar{z} \partial_{\bar{z}}+ w \partial_w + \bar{w} \partial_{\bar{w}}$.

It is worthy to disentangle the mixed variables  present in the last, ``drift" term, by repeating the Lamperti transformation defined by \eqref{lamperti} in the hermitian case (with $N$ replaced by $2N$ and $Q$ instead of $z$).  
This change of variables leads to the free diffusion
\begin{eqnarray}
	\partial_{\tau'} D' = \frac{1}{N} \partial_{w'\bar{w}'} D' .
\end{eqnarray}
We contrast this  equation to its hermitian counterpart \eqref{hermdiff}. Again, it is exactly integrable, and the case of free diffusion was considered by us in~\cite{USPRL,USNPB}. Note that this time the diffusion is two-dimensional, and the Laplace operator  acts in  the $w'$  space, which, in standard treatments, is largely ignored by serving merely as a regulator.  Alike in the hermitian case, we may suspect the emergence of the Burgers structure, provided we apply Cole-Hopf transformation. Since we have at our disposal two complex variables $z'$ and $w'$, we may perform two independent Cole-Hopf transformations
\be
g'\equiv\frac{1}{N}\del_{z'}\ln D', \quad 
u'\equiv\frac{1}{N}\del_{w'}\ln D',
\ee
which  satisfy~\cite{USPRL,USNPB}
\begin{eqnarray}
\del_{\tau'}g' &= \frac{1}{N}\del_{w'\bar{w'}}g'+\del_{z'}(u'\bar{u'}),  \\
\del_{\tau'}u' &= \frac{1}{N}\del_{w'\bar{w'}}u'+\del_{w'}(u'\bar{u'}) .
\label{quatBurg}
\end{eqnarray}
Let us then perform the macroscopic and microscopic limit of above equations. 

\subsection{Macroscopic limit} 
In the large $N$ limit, the second equation (\ref{quatBurg}) takes the form of an inviscid Burgers equation in 2+1 dimensions
\be
\partial_{\tau'}u'=\partial_{w'} |u'|^2.
\ee
In our case this equation can be simplified due to the rotational symmetry. From now on, we follow the solution presented in ~\cite{USPRL}, 
modulo the primed variables reflecting the Ornstein-Uhlenbeck process. Introducing $v'=|u'|$ and the radial variable $r'=|w'|$ we recover the Euler equation known from hydrodynamics.  The standard solution, using the method of characteristics, yields $v'=v_0'(r'+\tau'v')$, where the initial condition $X_0=0$ corresponds to $v_0'(r')=r'/(|z'|^2+r'^2)$. We identify the $v'^2$ with the eigenvector correlator $O(z',\tau')$, with an explicit solution
\be
O(z', \tau')=\frac{1}{\pi \tau'^2} (\tau'-|z'|^2) .
\ee
Having the solution for $v'^2$, we can turn back to the first equation  $\partial_{\tau'}g'=\partial_{z'}v'^2$. Elementary integration and initial conditions lead to $g'=\bar{z}'/\tau'$ which in turn gives the spectral density \be
	\rho(z',\tau') = \frac{1}{\pi \tau'} \theta(\tau' - |z'|^2) .
\ee  
We can now return to the unprimed variables, using the same Lamperti formulae (\ref{lamperti}) and perform the stationary limit $\tau \rightarrow \infty$. In the end, the eigenvector correlator and the spectral density read simply
\be
O(z,\bar{z})&=&\frac{4a^2}{\pi} \left (\frac{1}{2a}-|z| ^2 \right ), \\
\rho(z,\bar{z})&=& \frac{2a}{\pi} \theta \left (\frac{1}{2a}-|z|^2 \right ),
\label{finalmacro}
\ee 
which reproduce the known GE results for $a=1/2$.

It is amusing to note, that historically, the first equation for $O(z,\bar{z})$  in the Ginibre ensemble  was delivered by Chalker and Mehlig~\cite{CHALKERMEHLIG} more than three decades after the  result for the uniform spectral density $\rho(z,\bar{z})$,  originally obtained by Ginibre. 
In our ``turbulent" formulation,  at least in the large $N$ limit,  the equation for the eigenvector correlator is of primary importance, and the solution for the spectral density follows trivially from the knowledge of the eigenvector correlator.  This observation points at the crucial difference between the hermitian and non-hermitian random matrix models - whereas in the hermitian case the spectral properties are dominant and the eigenvectors decouple,  in the non-hermitian case the eigenvectors control the spectral evolution. Technically, this observation was missed in the literature because the analytic structure driven by the $w$ variable was overlooked. 

\subsection{Microscopic limit} 
We refer the detailed discussion to the published work~\cite{USNPB}.  We mention only, that since the equation for characteristic determinant is given as exactly integrable, 2+1 dimensional diffusion equation, valid for any $N$ and any initial condition, unraveling the universal behavior at the edge of the spectrum is a consequence of certain limiting procedures of the exact solution.
Interestingly, the pre-shock wave appears in the non-hermitian case in the eigenvector correlator, contrary to the appearance of the spectral pre-shock wave in eigenvalue spectrum in the case of the hermitian ensemble. Looking at the neighborhood of the shock by parameterizing the fluctuations in the vicinity of the boundary as $|z|-1=\eta N^{-1/2}$, we recover the known universal  result for the microscopic behavior 
\be
\rho(\eta) \approx \frac{1}{2\pi} {\rm Erfc} (\sqrt{2} \eta).
\ee
The form of the unfolding is expected from the general geometric argument  since the number of eigenvalues on the surface of the disc grows alike $N$, the scaling on the boundary has to grow alike $\sqrt{N}$. 
Note that contrary to the hermitian case,  the spectrum does not oscillate wildly at the edge,  but rather smoothly interpolates between  the plateau at $1/\pi$ and 0.
This  behavior can be linked to the fact, that the viscosity  in the non-hermitian case has a positive sign. 

\section{A qualitative relationship between the dynamics of eigenvectors and eigenvalues based on numerical experiments}
\begin{figure}[ht!]
	\centering
	\includegraphics[width=.7\textwidth]{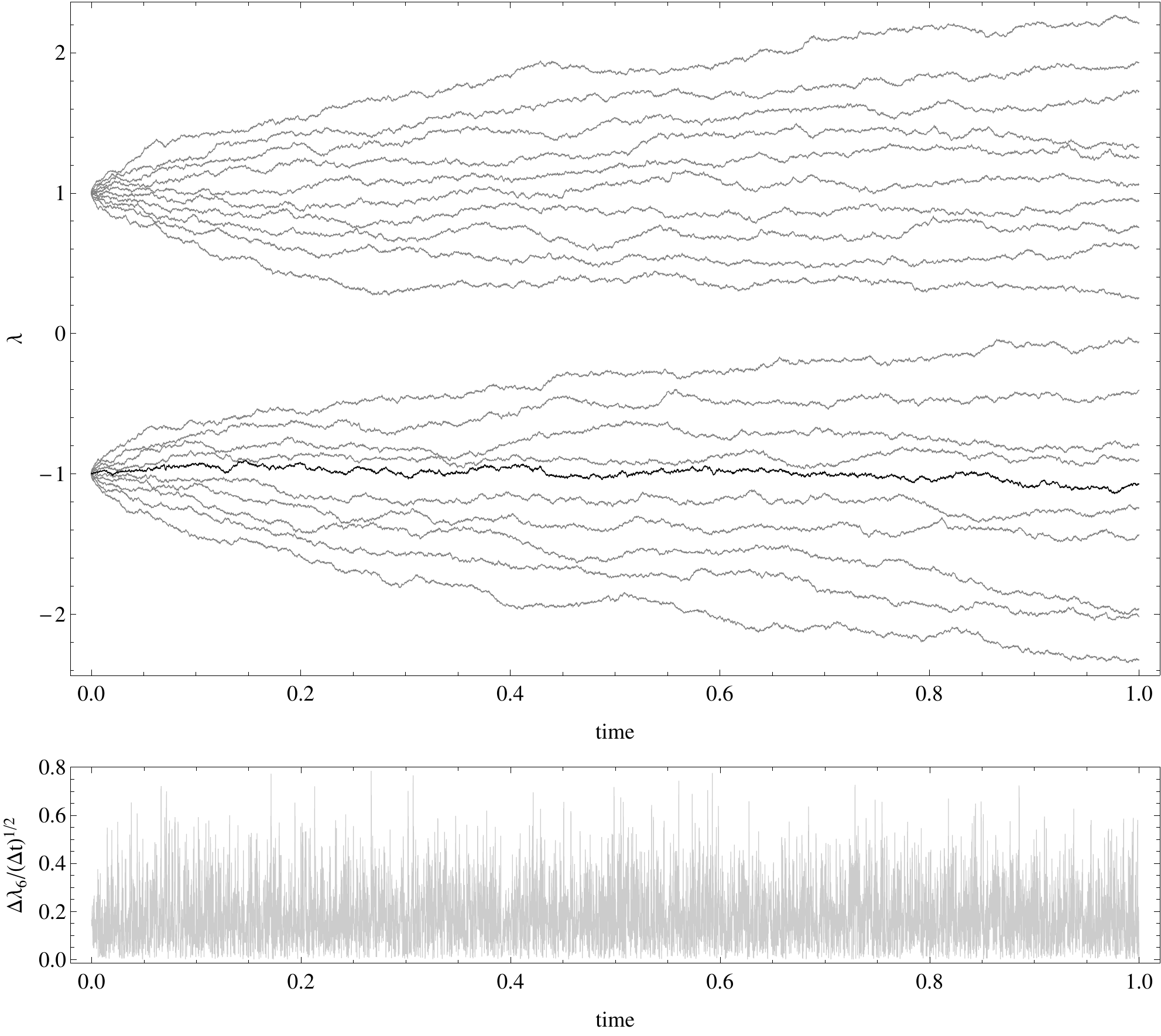}
      \caption{Upper inlet, a single numerical realization of stochastic behaviour of $N=20$ eigenvalues governed by eq. \eqref{LangGUE}. Initially the eigenvalues were put at $-1,1$ with equal multiplicities. Lower inlet, the jump amplitude of a given eigenvalue (in bold in the upper plot), normalized by the square root of the time step used in the simulation.}
      \label{guenum}
\end{figure}

To gain some more insight into the intertwined dynamics of the complex eigenvalues and eigenvectors of non-hermitian matrices, let us perform some numerical experiments.
First, for comparison, we turn to hermitian matrices. As was mentioned before, in this case, the diffusion (\ref{entrydiff}) induces the Langevin equation (\ref{LangGUE}) for the eigenvalues - no coupling to the eigenvectors is present. 
It is worth mentioning that the eigenvector dynamics \textit{does} depend on the eigenvalues. The associated stochastic equation of motion is given by:
\begin{equation}
\label{hev}
{\rm d} \ket{\psi_i(\tau)} = \frac{1}{\sqrt{N}} \sum_{j=1,j\neq i}^N \frac{{\rm d}B_{ij}(\tau)}{\lambda_i - \lambda_j} \ket{\psi_j} - \frac{1}{2N} \sum_{j=1, j\neq i}^N \frac{{\rm d}\tau}{(\lambda_i - \lambda_j)^2} \ket{\psi_i},
\end{equation}
and has been studied extensively in \cite{BOUCHAUD} (the $\ket{\psi_i}$ is an eigenvector corresponding to eigenvalue $\lambda_i$, ${\rm d}B_{ij}$ is a multi-dimensional Brownian motion). Nonetheless, the evolution of the eigenvalues does not depend on the eigenvectors and when interested only in the dynamics of $\lambda_i$, we can ignore the changes of $\ket{\psi_i}$'s. The resulting process is depicted in Fig.~\ref{guenum} where we present the eigenvalue trajectories of a $N=20$ hermitian matrix initiated with two distinct  eigenvalues $\lambda=-1,1$ with equal multiplicities. Additionally, we have computed the jump amplitude of a particular eigenvalue normalized by the simulations time step. Note that there is no distinct dependence of the jump  on how close the eigenvalue is to its neighbours. 

\begin{figure}[ht!]
	\centering
	\includegraphics[width=.8\textwidth]{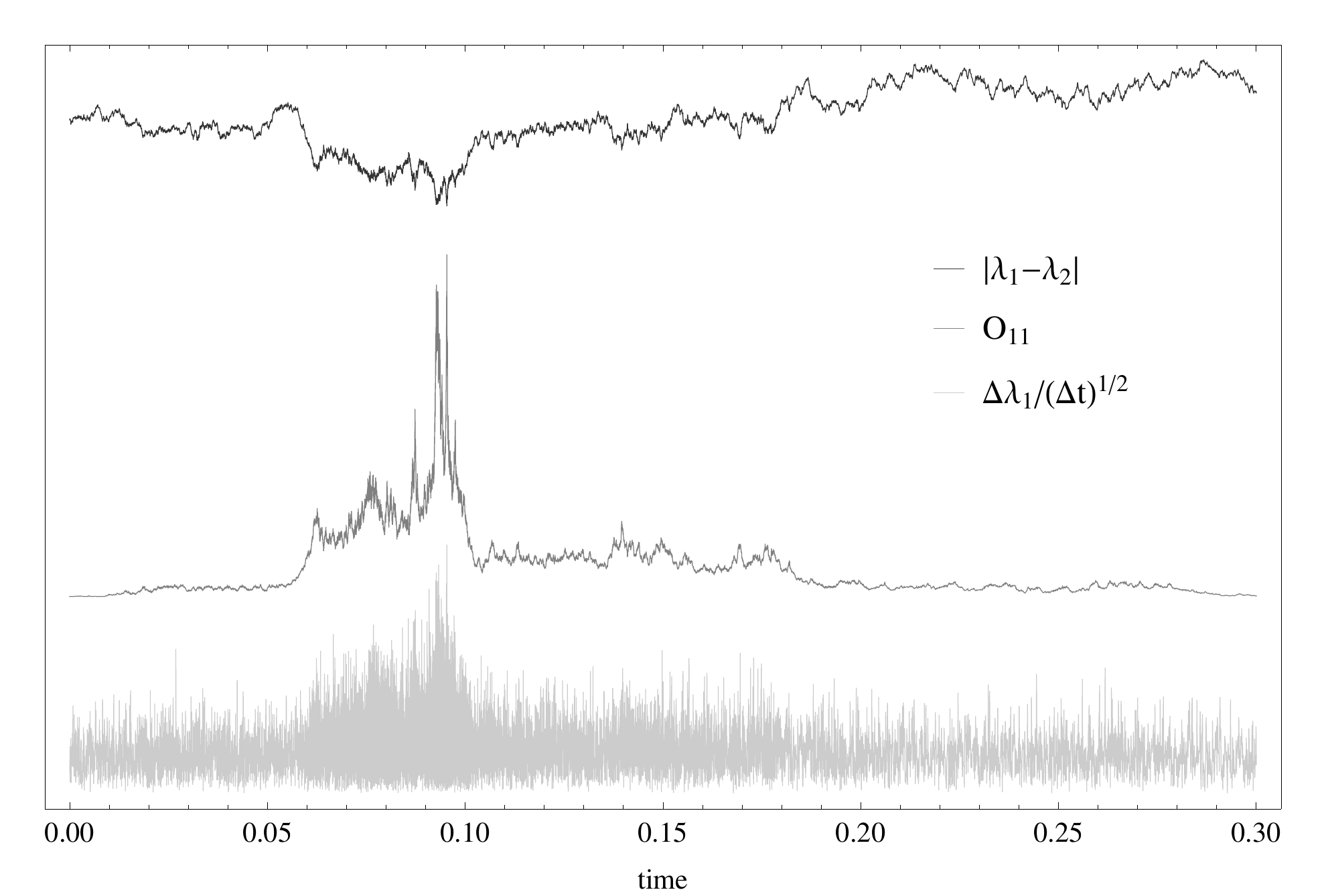}
      \caption{Time series of eigenvalue distance $|\lambda_1 -\lambda_2|$, eigenvector $O_{11}$ and eigenvalue diffusion distance $\delta \lambda_1 = \Delta \lambda_1/(\Delta t)^{1/2}$. Corresponding vertical axes are out of scale, we identify the time $t_c = 0.1$ to be of both minimal distance $|\lambda_1 - \lambda_2|$ and maximal values of $O_{11}$ and ${\delta \lambda_1}$.}
      \label{genum2}
\end{figure}

In the case of non-hermitian dynamics, the Langevin equations are not readily available - we relegate their derivation and study to future work. Nonetheless, as was argued in Sec.~\ref{secge}, we now know that the eigenvectors and in particular $O_{ii}$'s are crucial for the dynamics of complex eigenvalues. 
\begin{figure}[ht!]
	\centering
	\includegraphics[width=.8\textwidth]{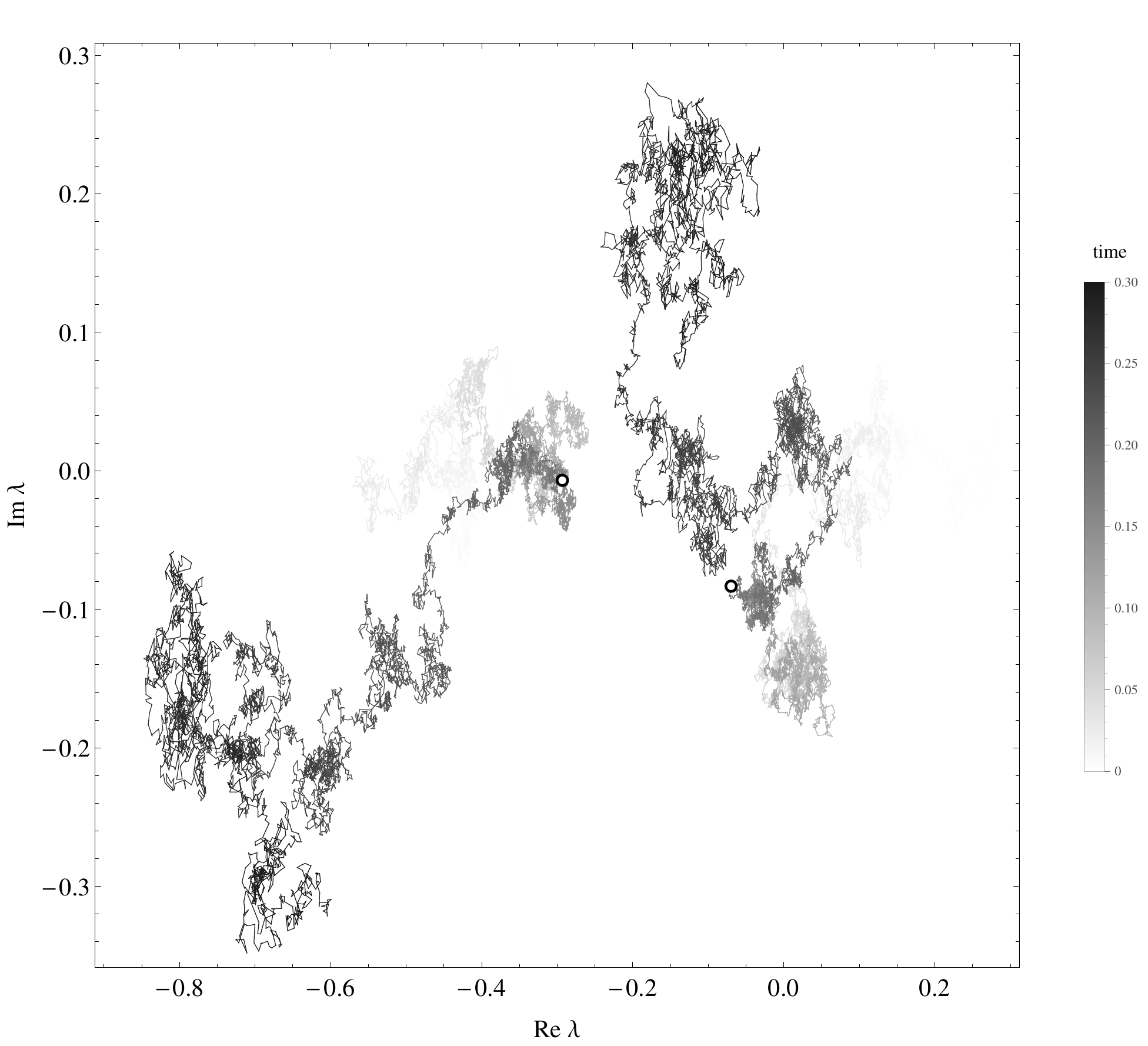}
      \caption{Numerical realization of a stochastic behaviour of $N=2$ eigenvalues of a non-hermitian matrix diffusing according to (\ref{nonhdiff}), with $a=0$. Initial conditions are $\lambda_{1}=0.3,\lambda_2=-0.3$ and the color of paths encode time evolution. Black edged white dots represent the position of $\lambda$'s at time $t=0.1$ when the distance is minimized (see Fig. \ref{genum2}).}
      \label{genum1}
\end{figure}
To show this, we focus on an example of a non-hermitian evolution of a $N=2$ matrix starting from ${\rm diag}(-0.3,0.3)$. In Fig.~\ref{guenum}, we observe the eigenvalues covering the complex plane in a diffusive manner. It is also expected that they repel each other. To perform a closer inspection (see Fig.~\ref{genum2}), we plot three characteristics of their dynamics - the distance between the eigenvalues $|\lambda_1 - \lambda_2|$, the eigenvector correlator $O_{11} = \braket{L_1|L_1} \braket{R_1|R_1}$ and the normalized jump $\Delta \lambda_1/(\Delta t)^{1/2}$ of the first eigenvalue, all as a function of time. We chose to ignore both $O_{22}=O_{11}$ and $O_{12}=1-O_{11}$ since they do not offer any additional information. The most interesting feature of this particular realization occurs around the time $t_c=0.1$ of minimal eigenvalue distance (this precise moment is depicted by white dots on Fig.~\ref{genum2}). We observe that as the distance gets smaller, the $O_{11}$ blows up in a correlated manner. This is accompanied by an increase in the jump amplitude of the eigenvalue. W have checked that this effect prevails when matrix size is larger than two. Note again that it was not present for eigenvalues of hermitian matrices, for which the distance between the eigenvalues also drives the evolution. We therefore consider this effect as a qualitative demonstration of the co-dependence between the evolutions of eigenvalues and eigenvectors in this scenario.

\section{Unexpected links}
Several unexpected links between the static hermitian and non-hermitian random matrix models were noted in the past~\cite{JAROSZNOWAK}. 
The spectrum  of hermitian matrices is real, but the main tool relies on introducing the complex valued resolvent (Green's function),  whose discontinuities allow to infer the spectral function, using the theory of analytic functions. In the large $N$ limit, a particular transform, known as the R-transform,  related to the Green's function by the functional inverse as  $R[G(z)] +1/G(z)=z$, plays the role of the analog of generating function of classical cumulants in the matrix-valued probability calculus. 
The R-transform constitutes the cornerstone of the free probability theory~\cite{VOICULESCU} and generates matrix-valued analogues of classical central limit theorems. In the case  of non-hermitian matrices, the spectrum is complex, but the  regulator $\epsilon^2$ in the  logarithmic potential (\ref{Coulomb})  behaves as the tip of an iceberg, pointing at a hidden algebraic structure. Indeed, in order to maintain the analogy to the hermitian case, one has to embed the  structure  of the generalized Green's functions in  the algebra of quaternions. In such a way, a second complex variable $w$, ``perpendicular" to $z$  emerges.  In the large $N$ limit, one can adapt  the Voiculescu construction for the R-transform by defining the quaternion valued  functional inverse ${\cal{R}}[{\cal{G}}(Q)]+1/{\cal{G}}(Q)=Q$ and thus allowing for non-hermitian  and non-commuting convolution of random matrices~\cite{JANIKNOWAK,JAROSZNOWAK}.
Surprisingly, the links between the hermitian and non-hermitian random matrix models stretch out  to the area of dynamic processes. In the case of the Gaussian randomness,  the exact diffusion  equation for the averaged characteristic polynomial finds its exact analogue for the averaged characteristic polynomial  valued in the algebra of quaternions. It turns out, that the ``hidden" variable $w$, ignored in standard treatment of non-hermitian random matrix models, plays a crucial role in determining the two-dimensional pattern of the spectral evolution.  In the large $N$
 limit, hermitian and non-hermitian Smoluchowski-Fokker-Planck equations take the surprisingly  similar form of a Burgers-like structure. 
 In general, the Voiculescu equation $\partial_{\tau} G +R(G)\partial_z G=0$ is  replaced  by its quaternionic counterpart~\cite{TARNOWSKI}

\be
\frac{\partial {\cal{G}}_{ab}}{\partial \tau}+ \sum\limits_{c,d=1}^{2}{\cal{R}}[{\cal{G}}]_{cd}\frac{\partial{\cal{G}}_{ab}}{\partial Q_{cd}}=0,
\ee
where latin indices label the two-by two quaternionic structure of $Q$, ${\cal G}$ and ${\cal{R}}$.  
In both cases, singularities  emerge.  However, in the hermitian case,  singularities appear in the flow of the eigenvalues, whereas in the case of non-hermitian ensembles, singularities appear in the flow of a certain correlator of left and right eigenvectors. In both cases, finite $N$ effects can be taken into account as an appearance of spectral viscosity proportional to $1/N$.  There is however a crucial difference in the sign - positive spectral viscosity smoothens the edge of the Ginibre spectrum, yielding universal  behavior given by the $\rm{Erfc}$ function, whereas negative spectral viscosity in the GUE triggers violent oscillations, leading to the formation of the so-called Airy kernel. 
Resolving the deep reasons for these links still remains one of the challenges of random matrix models. 
We summarize the unexpected links between the GUE and GE ensembles in  Table I.

\begin{table}
\begin{center}
 \begin{tabular}{||c c c ||} 
 \hline
  & GUE  & GE  \\ 
  \hline\hline
 Spectral density  & real  & complex  \\ 
 \hline
 Resolvent  & complex-valued & quaternion-valued  \\
  & $G(z)=\frac{1}{N} \left< {\rm Tr} (z-H)^{-1}  \right>$ \,\,\,\,\,\,\,&   ${\cal G}(Q)=\frac{1}{N} \left< {\rm bTr} (Q-{\cal X})^{-1}  \right>$\,\,\,\,\,\,\, \\
 \hline
 Determinant & $U(z, \tau) = \left < {\rm det} (z-H) \right >$\,\,\,\,\,\,\, & $D(Q, \tau) = \left < {\rm det} (Q-{\cal{X}}) \right >$ \,\,\,\,\,\,\,\\
 \hline
Diffusion eq.  & $\partial_{\tau} U= - \frac{1}{2N}\partial_{zz} U$\,\,\,\,\,\,\,\,   & $\partial_{\tau} D=  +\frac{1}{N}\partial_{w\bar{w}} D$\,\,\,\,\,\,\, \\ 
 \hline
 Viscosity &          negative           & positive \\
\hline 
Universal behavior   & oscillatory (Airy kernel) &  smooth  (Erfc) \\
\hline 
 R-transform & $R_{GUE} (G)=G $ &  ${\cal{R}}_{GG}({\cal{G}})= \left(    \begin{array}{cc} 0 & {\cal{G}}_{1\bar{1}}\\   
{\cal{G}}_{\bar{1}1} & 0 \end{array}  \right)$ \\
 \hline
 Voiculescu equation &  $ \frac{\partial G}{\partial \tau}+R(G)\frac{\partial G}{\partial z}=0$  & $
\frac{\partial {\cal{G}}_{ab}}{\partial \tau}+ \sum\limits_{c,d=1}^{2}{\cal{R}}[{\cal{G}}]_{cd}\frac{\partial{\cal{G}}_{ab}}{\partial Q_{cd}}=0
$ \\
 \hline 
 Pre-shock waves  &    Flow of eigenvalues    & Flow of eigenvector correlators \\
 \hline
 \hline 
\end{tabular}
\caption{Comparison of links between GUE and GE.}
\end{center}
\end{table}

\section{Conclusions}
The presented results borrow to a large extent from the conclusions obtained in the series of papers of the present authors~\cite{USOLD,USACTA,USPRL,USNPB}, but also include new solutions. 
First, we adapted the turbulent scenario to the Ornstein-Uhlenbeck process for GUE. Technical details are deferred to Appendix A. Then, by a set of transformations, we provided an exact mapping between the Ornstein-Uhlenbeck process and free diffusion. This mapping allowed us to interpolate smoothly between the microscopic limit (Dyson's local equilibrium) and the macroscopic limit (Dyson's global equilibrium). Second, we  have repeated the same scenario of the Ornstein-Uhlenbeck process for the Ginibre ensemble. Again, we relegate technical details to Appendix B. 
Last but not least, we tried to point at rather unexpected analogies and similarities in both examples.  We stressed that such analogies are detectable only when the quaternion variables are used. 

We have proven that a consistent description of non-hermitian ensembles require the knowledge of the detailed dynamics of co-evolving eigenvalues and eigenvectors. Moreover, at least in the large $N$ limit, the dynamics of eigenvectors plays a major role and leads directly to the inference of the spectral properties.  This is a dramatically different scenario comparing to the standard random matrix models, where the statistical properties of eigenvalues are of primary importance, and the properties of eigenvectors are basically trivial due to the their decoupling from the spectra and the fact that they are Haar distributed on $U(N)$.  We conjecture that the hidden dynamics of eigenvectors observed in the Ginibre ensemble, is a general feature of all non-hermitian random matrix models.

Our formalism could  be exploited to expand the area of application of non-hermitian random matrix ensembles within problems of growth, charged droplets in quantum Hall effect and gauge theory/geometry relations in string theory  beyond the subclass of complex matrices represented by normal matrices. 

One of the challenges is an explanation, why, despite being so different, the Smoluchowski-Fokker-Planck equations for hermitian and non-hermitian random matrix models exhibit structural similarity to simple models of turbulence, where so-called Burgers equation plays the vital role, establishing the flow of the spectral density of eigenvalues in the case of the hermitian or unitary ensembles  and the flow of certain eigenvector correlator in the case of non-hermitian ensembles. 
Another challenge relies in completing the Langevin-like equations (\ref{LangGUE}) adapted for non-hermitian cases. 

We believe that our findings will contribute to the understanding of several puzzles of non-hermitian dynamics, alike extreme sensitivity of spectra of non-hermitian systems to perturbations~\cite{CHALKERMEHLIG,SAVIN} and the sign problem of certain Euclidean Dirac operators.

\section*{Acknowledgments}
MAN  thanks the organizers of the 7th International Conference on Unsolved Problems on Noise, Barcelona, Spain, for hospitality. We appreciate collaboration with Zdzis\l{}aw Burda, Ewa Gudowska-Nowak,  Romuald Janik, Andrzej Jarosz, Jerzy Jurkiewicz, Gabor Papp and Ismail Zahed and  discussions with  Neil O'Connell, Yan Fyodorov, Mario Kieburg, Roger Tribe, Dima Savin and  Oleg Zaboronski.  This work was supported by the Grant DEC-2011/02/A/ST1/00119 of the National Centre of Science. WT appreciates partial support from the Ministry of Science and Higher Education through the Grant 0225/DIA/2015/44. 
%

\section*{Appendix A}
In what follows we derive the partial differential equation \eqref{eq:U}. For completeness, we again introduce the Ornstein-Uhlenbeck diffusion process in terms of a Smoluchowski-Fokker-Planck equation \eqref{jpdfeq}: 
\begin{eqnarray}
\label{jpdfeqdiff}
\del_\tau P(x,y,\tau)=\mathcal{A}(x,y) P(x,y,\tau),
\end{eqnarray}
with the Laplace operator
\begin{eqnarray}
 \mathcal{A}(x,y) & = \sum_{k=1}^N \left(\frac{1}{2N}\frac{\del^{2}}{\del x_{kk}^{2}}  +a\frac{\del}{\del x_{kk}}x_{kk}\right) +\frac{1}{4N}\sum_{i<j=1}^N \left ( \frac{\del^{2}}{\del x_{ij}^{2}}+ \frac{\del^{2}}{\del y_{ij}^{2}} \right ) + \nonumber \\ 
& + a\sum_{i<j=1}^N \left ( \frac{\del}{\del x_{ij}}x_{ij}+ \frac{\del}{\del
y_{ij}}y_{ij} \right ) . \label{prob1}
\end{eqnarray}
As a first step, we write the determinant as a Gaussian integral over a set of Grassmann variables $\eta_i, \bar \eta_i$:
\begin{eqnarray}
{\rm det}~{A} = C {\int {{\prod _{i,{j}=1}^N{{\rm d}{\eta }_{{i}}}}{{\rm d}{{{\overline\eta } }_{{j}}}}}}\,{\exp{\left({\overline{{\eta }_{{i}}}{A}_{{ij}}{\eta
}_{{j}}}\right)}}.\label{gras}
\end{eqnarray}
with an irrelevant proportionality constant $C$. This allows us to express the characteristic polynomial $U$ defined by \eqref{defu} in the following way:
\begin{eqnarray}
U(z,t) & = C \int \mathcal{D}[\bar{\eta},\eta,x,y]P(x,y,\tau)\exp{\left[\sum_{i,j=1}^N \bar{\eta}_i\left(z\delta_{ij}-H_{ij}\right)\eta_{j}\right]},\label{repr}
\end{eqnarray}
where the integration measure is defined by
\begin{eqnarray}
\mathcal{D}[\bar{\eta},\eta,x,y] & \equiv \prod _{i,j=1}^N{\rm d}\eta_{i}{\rm d}\overline{\eta}_{j} \prod_{k=1}^N{\rm d}x_{kk}\prod_{n<m=1}^N{\rm d}x_{nm}{\rm
d}{y}_{nm}.
\end{eqnarray}
The hermiticity condition ($H_{ij}=\bar{H}_{ji}$) is used to write the argument of the exponent of (\ref{repr}) in an explicit form:
\begin{eqnarray}
& T_g(\bar{\eta},\eta,x,y,z) \equiv \sum_{i,j=1}^N \bar{\eta}_i\left(z\delta_{ij}-H_{ij}\right)\eta_{j} = \sum_{r=1}^N\bar{\eta}_{r}\left(z-x_{rr}\right)\eta_{r} + \nonumber \\
& -\sum_{n<m=1}^N\left[
x_{nm}\left(\bar{\eta}_{n}\eta_{m}+\bar{\eta}_{m}\eta_{n}\right)+iy_{nm}\left(\bar{\eta}_{n}\eta_{m}-\bar{\eta}_{m}\eta_{n}\right)\right] . 
\end{eqnarray}
Now we make use of the diffusion equation \eqref{jpdfeqdiff} which, after integrating by parts, gives:
\begin{eqnarray}
\label{parts1}
	\del_\tau U = \int \mathcal{D} \partial_\tau P \exp \left ( T_g \right ) = \int \mathcal{D} \mathcal{A} P \exp \left ( T_g \right ) = \int \mathcal{D} P \tilde{\mathcal{A}} \exp \left ( T_g \right ),
\end{eqnarray}
with $\tilde{\mathcal{A}} = \mathcal{A} (a \to -a)$. We calculate the expression:
\be\nonumber
	\tilde{\mathcal{A}} \exp \left ( T_g \right )= & \left [ a \sum_{k=1}^N x_{kk}\bar{\eta}_k \eta_k+ \frac{1}{N} \sum_{i<j=1}^N \bar{\eta}_i \eta_j \bar{\eta}_j \eta_i  \right ] \exp \left ( T_g \right ) + \\ 
 & + \left [ a  \sum_{i<j=1}^N \left [ (x_{ij} + i y_{ij})\bar{\eta}_i \eta_j + (x_{ij}-i y_{ij}) \bar{\eta}_j \eta_i  \right ]\right ]  \exp \left ( T_g \right ) .
  \label{expr1}
\ee
by schematically writing down the action of derivatives on $\exp ( T_g )$:
\begin{eqnarray*}
\partial_{x_{kk}} & \quad \rightarrow \quad -\bar{\eta}_k \eta_k , \qquad\partial^2_{x_{kk}} \quad \rightarrow \quad 0 ,
\end{eqnarray*}
and for $i\ne j$:
\begin{eqnarray*}
\partial_{x_{ij}} & \quad \rightarrow \quad -(\bar{\eta}_i \eta_j + \bar{\eta}_j \eta_i ), \\
\partial_{y_{ij}} & \quad \rightarrow \quad -i(\bar{\eta}_i \eta_j - \bar{\eta}_j \eta_i ), \\
	\partial^2_{x_{ij}} & \quad \rightarrow \quad (\bar{\eta}_i \eta_j + \bar{\eta}_j \eta_i )(\bar{\eta}_i \eta_j + \bar{\eta}_j \eta_i ) =  -2\bar{\eta}_i \eta_j \bar{\eta}_j \eta_i, \\
	\partial^2_{y_{ij}} & \quad \rightarrow \quad -(\bar{\eta}_i \eta_j - \bar{\eta}_j \eta_i )(\bar{\eta}_i \eta_j - \bar{\eta}_j \eta_i ) = - 2\bar{\eta}_i \eta_j \bar{\eta}_j \eta_i. 
\end{eqnarray*}
We rewrite the terms of \eqref{expr1} accordingly:
\begin{eqnarray*}
& \sum_{i=1}^N\bar{\eta}_i\eta_i\exp\left(T_g\right)=\partial_z\exp\left(T_g\right), \\
& \sum_{i<j=1}^N\bar{\eta}_i\eta_i\bar{\eta}_j\eta_j\exp\left(T_g\right)=\frac 12\partial_{zz}\exp\left(T_g\right),\\ 
& 	\sum_{i=1}^N \bar{\eta}_i \partial_{\bar{\eta}_i} \exp \left ( T_g \right )  = \left [ z\del_z - \sum_{i=1}^N x_{ii}\bar{\eta}_i \eta_i - \sum_{i<j=1}^N \left [ (x_{ij} + i y_{ij})\bar{\eta}_i \eta_j + (x_{ij}-i y_{ij}) \bar{\eta}_j \eta_i  \right ] \right ] \exp \left ( T_g \right ).	
\end{eqnarray*}
We obtain thus, after recalling \eqref{parts1} and \eqref{expr1}:
\begin{eqnarray}
	\del_\tau U = \int \mathcal{D} P \left (- \frac{1}{2N} \del_{zz} +a z \del_z-  a  \sum_{i=1}^N \bar{\eta}_i \partial_{\bar{\eta}_i}  \right ) \exp \left ( T_g \right ),
\end{eqnarray}
where the last term is explicitly calculable upon integrating by parts
\begin{eqnarray}
\int \mathcal{D} P \sum_{i=1}^N \bar{\eta}_i \partial_{\bar{\eta}_i}  \exp \left ( T_g \right ) = N \int \mathcal{D} P \exp \left ( T_g \right ),
\end{eqnarray}
so that we arrive at the equation \eqref{eq:U}
\begin{eqnarray}
\del_{\tau}U(z,t)=-\frac{1}{2N}\del_{zz}U(z,\tau)+a z \del_{z}U(z,\tau)- a N U(z,\tau).
\end{eqnarray}


\section*{Appendix B} 
Here we present the derivation for the evolution equation \eqref{ggdiff}. The Ornstein-Uhlenbeck process in the non-hermitian case is given by Smoluchowski-Fokker-Planck equation for the jpdf $P(x,y,\tau)$:
\begin{eqnarray}
\del_\tau P(x,y,\tau)=\mathcal{B}(x,y) P(x,y,\tau),
\end{eqnarray}
with the operator
\begin{eqnarray}
\mathcal{B} = \frac{1}{4N} \sum_{i,j=1}^N \left ( \partial^2_{x_{ij}} + \partial^2_{y_{ij}} \right ) +  a  \sum_{i,j=1}^N \left ( \partial_{x_{ij}} x_{ij} + \partial_{y_{ij}} y_{ij} \right).
\end{eqnarray}
To proceed, we open the determinant defined in \eqref{cd} with the help of Grassmann variables $\eta_i,\xi_i$:
\begin{eqnarray*}
	D = C'\int \mathcal{D}[...] P(x,y,\tau) \exp \left [ \left ( \begin{array}{cc} \bar{\eta} & \bar{\xi} \end{array}\right ) \left ( \begin{array}{cc} z-X & -\bar{w} \\ w & \bar{z} - X^\dagger \end{array} \right ) \left ( \begin{array}{c} \eta \\ \xi \end{array} \right ) \right ] ,
\end{eqnarray*}
with $X_{ij} = x_{ij} + i y_{ij}, X^\dagger_{ij} = x_{ji} - i y_{ji}$, an irrelevant constant $C'$ and the joint measure 
\be
\qquad \mathcal{D}[\bar{\eta},\eta,\bar{\xi},\xi,x,y] & = \prod _{i=1}^N\dd \eta_{i}\dd \bar{\eta}_{i} \dd \xi_i \dd \bar{\xi}_i \prod_{i,j=1}^N\dd x_{ij}\dd y_{ij} .
\ee
The argument of the exponent is equal to
\begin{eqnarray*}
	S_g & \equiv  \sum_{i,j=1}^N \left [ \bar{\eta}_i (z-X)_{ij} \eta_j + \bar{\xi}_i (\bar{z} - X^\dagger)_{ij} \xi_j + w \bar{\xi}_i \eta_i -\bar{w} \bar{\eta}_i \xi_i \right ] = \\
	& = \sum_{i,j=1}^N \left [- x_{ij} (\bar{\eta}_i \eta_j + \bar{\xi}_j \xi_i ) -i y_{ij} \left ( \bar{\eta}_i \eta_j - \bar{\xi}_j \xi_i \right ) + z \bar{\eta}_i \eta_i + \bar{z} \bar{\xi}_i \xi_i + w \bar{\xi}_i \eta_i -\bar{w} \bar{\eta}_i \xi_i \right ]. 
\end{eqnarray*}
We make use of the diffusion equation SFP equation and integrate by parts:
\begin{eqnarray}
\label{part2}
	\del_\tau D = \int \mathcal{D} \partial_\tau P \exp \left ( S_g \right ) = \int \mathcal{D} \mathcal{B} P \exp \left ( S_g \right ) = \int \mathcal{D} P \tilde{\mathcal{B}} \exp (S_g),
\end{eqnarray}
where $\tilde{\mathcal{B}} = \mathcal{B} (a \to -a)$. The last integrand reads:
\begin{eqnarray}
\label{B}
	\tilde{\mathcal{B}} \exp (S_g) = \left [ \frac{1}{N} \sum_{i,j=1}^N \bar{\eta}_i \eta_j \bar{\xi}_j \xi_i +  a  \sum_{i,j=1}^N \left ( (x_{ij} + i y_{ij})\bar{\eta}_i \eta_j + (x_{ij}-i y_{ij}) \bar{\xi}_j \xi_i  \right ) \right ] \exp (S_g) , \qquad )
\end{eqnarray}
where we used the schematic formulas acting on $\exp (S_g)$:
\begin{eqnarray*}
\partial_{x_{ij}} & \quad \rightarrow \quad -(\bar{\eta}_i \eta_j + \bar{\xi}_j \xi_i ), \\
\partial_{y_{ij}} & \quad \rightarrow \quad -i(\bar{\eta}_i \eta_j - \bar{\xi}_j \xi_i ), \\
	\partial^2_{x_{ij}} & \quad \rightarrow \quad (\bar{\eta}_i \eta_j + \bar{\xi}_j \xi_i )(\bar{\eta}_i \eta_j + \bar{\xi}_j \xi_i ) = 2 \bar{\eta}_i \eta_j \bar{\xi}_j \xi_i, \\
	\partial^2_{y_{ij}} & \quad \rightarrow \quad -(\bar{\eta}_i \eta_j - \bar{\xi}_j \xi_i )(\bar{\eta}_i \eta_j - \bar{\xi}_j \xi_i ) = 2 \bar{\eta}_i \eta_j \bar{\xi}_j \xi_i .
\end{eqnarray*}
To continue, we rewrite the terms of \eqref{B} proportional to $a$ as follows:
\begin{eqnarray}
 &\sum_{i=1}^N \bar{\eta}_i \eta_i \exp (S_g) = \partial_z \exp (S_g), \qquad \sum_{i=1}^N \bar{\xi}_i \xi_i \exp (S_g) = \partial_{\bar{z}} \exp (S_g), \nonumber \\
  & \sum_{i=1}^N \bar{\eta}_i \xi_i \exp (S_g) = - \partial_{\bar{w}} \exp (S_g), \qquad \sum_{i=1}^N \bar{\xi}_i \eta_i \exp (S_g) = \partial_{w} \exp (S_g), \nonumber \\
	& \sum_{i=1}^N \bar{\eta}_i \partial_{\bar{\eta}_i} \exp (S_g) = \left ( - \sum_{i,j=1}^N (x_{ij}+i y_{ij}) \bar{\eta}_i \eta_j + z \sum_{i=1}^N \bar{\eta}_i \eta_i - \bar{w} \sum_{i=1}^N \bar{\eta}_i \xi_i \right ) \exp (S_g), \nonumber \\
	& \sum_{j=1}^N \bar{\xi}_j \partial_{\bar{\xi}_j} \exp (S_g) = \left ( -\sum_{i,j=1}^N (x_{ij}-i y_{ij}) \bar{\xi}_j \xi_i + \sum_{j=1}^N \bar{z} \bar{\xi}_j \xi_j + w \sum_{j=1}^N \bar{\xi}_j \eta_j \right ) \exp (S_g), \nonumber
\end{eqnarray}
so that
\begin{eqnarray}
	 & a \sum_{i,j=1}^N \left ( (x_{ij} + i y_{ij})\bar{\eta}_i \eta_j + (x_{ij}-i y_{ij}) \bar{\xi}_j \xi_i  \right ) \exp (S_g) = \nonumber \\
	 & = \left [ a  \sum_{i=1}^N \left ( - \bar{\eta}_i \partial_{\bar{\eta}_i} - \bar{\xi}_j \partial_{\bar{\xi}_j} \right ) +  a (z \partial_z+\bar{z} \partial_{\bar{z}}+ w \partial_w + \bar{w} \partial_{\bar{w}})  \right ] \exp (S_g).
\end{eqnarray}
Plugging the above expressions into \eqref{part2}, gives:
\begin{eqnarray}
	\del_\tau D = \int \mathcal{D} P & \left ( \frac{1}{N} \sum_{i,j=1}^N \bar{\eta}_i \eta_j \bar{\xi}_j \xi_i -  a  \sum_{i=1}^N \left ( \bar{\eta}_i \partial_{\bar{\eta}_i} + \bar{\xi}_j \partial_{\bar{\xi}_j} \right ) + \right .\nonumber \\
	& +  a  (z \partial_z+\bar{z} \partial_{\bar{z}}+ w \partial_w + \bar{w} \partial_{\bar{w}}) \Bigg ) \exp (S_g),
\end{eqnarray}
where the first term is expressible as
\begin{eqnarray}
	\partial_{w\bar{w}} D & = \int \mathcal{D} P \sum_{i,j=1}^N \bar{\eta}_j \eta_i \bar{\xi}_i \xi_j \exp \left ( S_g \right ),
\end{eqnarray}
and the second one reads
\begin{eqnarray}
\int \mathcal{D} P \sum_{i=1}^N \left ( \bar{\eta}_i \partial_{\bar{\eta}_i} + \bar{\xi}_j \partial_{\bar{\xi}_j} \right ) \exp \left ( S_g \right ) = 2N \int \mathcal{D} P \exp \left ( S_g \right ).
\end{eqnarray}
Taking them into account we finally obtain equation \eqref{ggdiff}
\begin{eqnarray}
	\del_\tau D = \frac{1}{N} \del_{w\bar{w}} D - 2N  a  D +  a  \textrm{d} D,
\end{eqnarray}
where $\textrm{d} = z \partial_z+\bar{z} \partial_{\bar{z}}+ w \partial_w + \bar{w} \partial_{\bar{w}}$.



\end{document}